# Preferential Forest-assembly of Single-Wall Carbon Nanotubes on Low-energy Electron-beam Patterned Nafion Films

*Haoyan Wei‡, Sang Nyon Kim†, Harris L. Marcus‡\*, and Fotios Papadimitrakopoulos†\**

‡ Department of Materials Science and Engineering, Institute of Materials Science, University of Connecticut, Storrs, CT 06269, USA

† Nanomaterials Optoelectronics Laboratory, Polymer Program, Institute of Materials Science, Department of Chemistry, University of Connecticut, Storrs, CT 06269, USA

Email Contact: Haoyan Wei, wtaurus@msn.com

Abstract

With the aid of low-energy (500 eV) electron-beam direct writing, patterns of perpendicularly-aligned Single-wall carbon nanotube (SWNT) forests were realized on Nafion modified substrates via $Fe^{3+}$-assisted self-assembly. Infrared spectroscopy (IR), atomic force microscopy (AFM) profilometry and contact angle measurements indicated that low-energy electron-beam cleaved the hydrophilic side chains ($-SO_3H$ and C-O-C) of Nafion to low molecular byproducts that sublimed in the ultra-high vacuum (UHV) environment exposing the hydrophobic Nafion backbone. Auger mapping and AFM microscopy affirmed that the exposed hydrophobic domains absorbed considerably less $Fe^{3+}$ ions upon exposure to pH 2.2 aqueous $FeCl_3$ solution, which yield considerably less FeO(OH)/FeOCl precipitates (FeO(OH) in majority) upon washing with lightly basic DMF solution containing trace amounts of adsorbed moisture. Such differential deposition of FeO(OH)/FeOCl precipitates provided the basis for the patterned site-specific self-assembly of SWNT forests as demonstrated by AFM and resonance Raman spectroscopy.

# 1. Introduction

Single- and multi-wall carbon nanotubes (SWNTs, MWNTs) have attracted considerable attention due to their unique structures, remarkable mechanical and electrical properties, as well as chemical stability[1-4]. SWNTs have found applicability to a wide range of electronics applications including nanodevices[5-9], sensors[10, 11] and field emitters[12, 13]. Their unique one-dimensional electronic structures make them ideal molecular wires to transport electrons from and to biological entities, such as peptides[14-18], bound along their length or on their carboxylated ends[19] and therefore have received considerable attention in designing nano-scale biosensors. The unique electron emission properties of vertically oriented nanotubes have attracted the interest in the industry such as flat panel display[13, 20], parallel electron-beam nanolithography and miniature X-ray generators[21]. For a number of these applications, it is very important to be able to produce uniform carbon nanotubes with controlled orientation normal to the substrates. In addition, in order to achieve point-source emission and localized electrochemical sensing, formation of spaced and laterally patterned SWNT probes with insulated side walls is also desired[22]. A number of researchers have reported on the growth of SWNTs and MWNTs using chemical vapor deposition (CVD) on patterned catalyst surfaces achieved via various methods such as standard lithography[23, 24], soft lithography[25], ink-jet printing[26] and nano-channel templates[27, 28]. Although the CVD-grown carbon nanotubes possess the right orientation to serve as probes, they are loosely packed (density around $10^{11}$ cm$^{-2}$)[28], which renders them extremely difficult to be handled in the presence of solvents and upon drying they easily collapse. The co-existence of various types of SWNTs (metallic (*met-*) and semiconducting (*sem-*)) could also pose serious problems for electronic devices in case only semiconducting or only metallic nanoprobes are needed[20].

Prior research in our laboratory have shown that rope-lattice dense SWNT forests (density on the order of $10^{13}$ cm$^{-2}$) can be readily obtained by assembling nanotubes from dimethylformamide (DMF) dispersion onto an underlying $Fe^{3+}$/Nafion composite bed[29]. On the basis of this development, and the above concerns, we have directed our patterning efforts on self-assembled nanotube forests, which can also take advantage of post-synthesis SWNTs separation according to length[30] and type (*sem-* versus

*met*-)[31]. Similar SWNT forest organization have been more recently reported using thiol functionalization of SWNT[32-34] as well as microcontact printing of SWNT dot arrays[35]. In contrast to $Fe^{3+}$-assisted forest-assembly, these require longer adsorption time[33-35] and attain lower surface coverage[32].

In the present contribution, we are reporting our initial findings in patterning these SWNT forests using low-energy electron-beam (500 eV) direct writing. The choice of low-energy electron-beam patterning was adopted for the following reasons: (1) The mean free path of low-energy electrons (10~500 eV) is small[36], thus most of the energy will be deposited onto the top surface of polymer films[37-39]. In addition, maximum cross section occurs at energy 2~3 times the binding energy for the given elements[40] (carbon 1s 284.6 eV, sulfur 2p 164.05 eV and sulfur 2s 229.2 eV) which are the main constituents of Nafion we are interested in. (2) Present e-beam lithography can still work very well in the nano-scale range at this voltage. (3) Sub-1keV is required to avoid electrical arcing in parallel electron column arrays which can increase the throughput substantially[37]. (4) Potential cross-contamination caused by chemical processing in traditional lithography can be prevented since no direct substrate contact is involved. We currently demonstrate that exposure of Nafion to moderate dose (c.a. 250~1250 μC/cm$^2$) of low-energy electrons (500 eV) causes a substantial loss of the hydrophilic side chains of Nafion on the immediate proximity of the surface of Nafion films leaving behind a relatively hydrophobic surface, which impedes the absorption of $Fe^{3+}$ ions and results in reduced assembly of SWNT forest arrays to the exposed regions.

## 2. Experimental Section

5% Nafion was purchased from Aldrich in water and lower aliphatic alcohols (1100 equivalent weight; *i.e.*, 1100 g of polymer per mole of −SO$_3$H groups). Iron(III) chloride hexahydrate (FeCl$_3$•6H$_2$O, A.C.S. reagent), nitric acid (98%), sulfuric acid (96.4%) and hydrochloric acid (38%) were also obtained from Aldrich and used as is. A.C.S. reagent dimethylformamide (DMF) was purchased from J.T.Baker. Hydrogen peroxide (30%) and A.C.S. certified methanol were obtained from Fisher Scientific. Millipore quality deionized water with resistivity larger than 18 MΩ was used for all experiments. Si(100) wafers were obtained from Montco Silicon and cleaned in Pirhana solution (concentrated $H_2SO_4$ and 30% $H_2O_2$, 7:3 v/v) at 90°C for 30 min. HiPco SWNTs were purchased from Carbon Nanotechnologies, Inc. Following the previously established protocol[19, 41, 42], pristine SWNTs were treated in 3:1 mixture of $HNO_3$ and $H_2SO_4$ with sonication for 4 hrs at 70°C, filtered, washed with copious deionized water until the pH of filtrated water reaches neutral and dried overnight in vacuum. Sonicating these purified shortened-SWNTs (s-SWNTs) in DMF resulted in stable dispersion.

SWNT forests were fabricated using a procedure described previously[29, 43]. Si substrates were first modified with Nafion via dipping in 1 mg/ml Nafion solution in methanol/$H_2O$ (9:1 v/v) for 30 min to form a smooth and uniformly negatively charged surface. SWNT/$Fe^{3+}$ assemblies were then obtained by sequential dipping of substrates in (i) $FeCl_3$ (pH 2.2, 15 min) solution, (ii) a quick wash in aqueous HCl solution (pH < 4) to remove loosely bound $Fe^{3+}$ ions, (iii) brief wash in DMF (pH 12.7) to remove excess water and facilitate $Fe^{3+}$ ions to transform into their basic hydroxide form[43], and (iv) a 30 min immersion in DMF dispersed SWNTs (pH 8.5) to enable the assembly of SWNT forest arrays.

Low-energy (500 eV) electron-beam patterning of Nafion modified Si substrates was performed in the PHI Auger 590 system with chamber pressure $10^{-9}$-$10^{-10}$ torr through a 500 mesh Transmission Electron Microscope (TEM) copper grid blocking electrons from areas where the subsequent SWNT/$Fe^{3+}$ deposition is desired (Scheme 1). A Faraday cup was used to measure the electron-beam current to calculate the doses for exposure. Infrared Spectra (IR) samples were obtained on Nafion 112 membranes (50 μm thick, made from DuPont) with 2.5 x 3.5 mm irradiated rectangles at elevated accelerating voltage 3 keV to obtain sufficient penetration depth for detection.

AFM characterization was done on Topometrix Explorer and Asylum Research MFP-3D in both contact mode and AC mode (Tapping Mode) to investigate the exposed Nafion films, iron deposits, and SWNT patterns. Liquid drop contact angle measurements on pristine and irradiated Nafion films were performed with water using the sessile drop mode (ramé-hart goniometer Model 100). IR Attenuated Total Reflection (ATR) was done on Spectra Tech IR-PLAN to study the possible irradiation mechanism on Nafion 112 membranes using a Germanium (Ge) crystal with 1 μm probing depth.

Spectra were collected before and after e-beam irradiation and plotted without normalization. XPS survey of dipping acquired Nafion films was made on Perkin-Elmer/PHI multi-probe surface analysis system with dual anode nonmonochromatic X-ray sources (PHI model 04-548) operated at 15 keV and 250 W to produce Al X-rays. The multi-probe system was maintained at a base pressure of c.a. $10^{-9}$ torr. XPS data were collected at pass energy of 89.45 eV. Controlled deposition of iron hydroxides was mapped in the same system by PHI Auger 595 at 3 keV collecting elemental Fe LMM (atomic shells) at peak energy 598 eV. Morphology and crystallinity of iron deposits precipitates were investigated using a JEOL 2010 TEM operating at 200 keV accelerating voltage. Preferential self-assembly of SWNT forest was characterized by both AFM and Raman resonance spectroscopy using the AFM instruments mentioned above and a Renishaw Ramanscope 2000 equipped with a 785 nm laser source focused on a 1 μm spot by a 100x objective lens respectively.

## 3. Results and Discussion

Depending on the dielectric constant ($\varepsilon$) of the organic solvent used for Nafion, the mixture can be categorized into three states (1) solution ($\varepsilon>10$), (2) colloid ($3<\varepsilon<10$), and (3) precipitates ($\varepsilon<3$)[44]. Based on the solvent mixture (9/1 methanol/water) used in this study ($\varepsilon>38$), Nafion is in the form of clear solution in micellar conformation (hydrodynamic radius in the order of c.a. 100 nm)[43] with most of its hydrophobic polytetrafluoroethylene (PTFE) backbone buried inside and the hydrophilic sulfonate groups located outside of the micelle[44]. Upon immersion into the acidic Nafion solution (pH~3), the partially protonated silanol groups (Si-OH) attracted the negatively charged Nafion micelles to deposit onto the Si substrate. This dipping process were shown to result in a flattened or pancale-like Nafion assemblies with thickness in the order of 10 nm[43]. This rendered the top surface of this assembly extremely hydrophilic (contact angle < 10°). Successful deposition of Nafion film was further confirmed by Auger investigation as shown in Figure S1 in supplemental information. Nafion modified substrates were immediately transferred to an Auger system chamber for electron-beam patterning through a TEM grid (Scheme 1). Exposed Nafion samples were examined with AFM, both surface topology and friction force in contact mode (Figure 1), clear patterns were observed in both imaging modes. The topological depression of irradiated regions, as shown in Figure 1a & c, indicated that mass loss occurred during electron-beam exposure. The fact that incident electrons impinged the sample surface at 45° instead of 90° (standard configuration for Auger system) made it difficult to determine the precise height difference as illustrated by the two orthogonal scans (Figure 1c), in which the measured values varied from 2 to 6 nm. Figure 1b gives the lateral force microscopy image which mapped surface chemical functionality[45]. Distinct friction-force differences were detected between electron-beam irradiated areas (squares) and unirradiated regions (grids), indicating an underlying transformation of Nafion's surface chemistry upon exposure to energetic electrons.

In order to obtain a better insight to Nafion's chemical transformation upon exposure to low-energy electron-beam irradiation, FTIR-ATR was utilized, which gives detailed information on the chemical functionalities. For our FTIR-ATR setup, Nafion film assembled according to methodology described above produced insufficient signal-to-noise data and therefore we resorted to thicker Nafion membranes (c.a. 50 μm). Since typical ATR detection depth is in the order of 1 μm, these thicker Nafion membranes were exposed to elevated beam voltage (3 keV) in order to increase the interaction depth with the e-beam. In Figure 2, the decreases of IR signal intensity for both $SO_3^-$ symmetric stretching mode (peak centered at 1060 $cm^{-1}$) and C-O-C symmetric stretching mode (at 982 $cm^{-1}$ and 970 $cm^{-1}$) were comparable for electron doses 250 μC/$cm^2$ and above. This is attributed to the small difference of dissociation energy between S-C and C-O bonds (280 kJ/mol and 355 kJ/mol respectively). This promotes the preferential scission of Nafion's side chains that could occur at three potential positions[46] indicated by the dashed lines on Nafion's structures (inset in Figure 2). We presently believe that for the irradiated section the majority of Nafion side chains located in the proximity of the surface were cleaved off and due to their low molecular weight these groups were sublimed to the ultra high vacuum ($10^{-9}$-$10^{-10}$ torr). The scission of hydrophilic side chains, particularly the anionic $SO_3^-$ groups, is expected to significantly reduce the ability of irradiated Nafion to withhold metal cations (*i.e.* $Fe^{3+}$). The fact that these three IR bands still maintained considerable intensity after e-beam exposure to irradiation dosages as high as 1250 μC/$cm^2$ originates from the fact that the penetration depth of 3 keV electron-beam is much shallower (for reference, electron range with same energy for PMMA is around 200~300 nm[47]) than the probing depth of the ATR equipped with Ge crystal (1 μm).

Hobson et al.[46, 48, 49] reported that upon e-beam irradiation in the presence of oxygen (200 ppm oxygen under atmospheric conditions), the formation of carboxyl functionalities take place at the sites of scission. This was witnessed by the emergence of the C=O stretch at 1770 cm$^{-1}$, which was assigned to the formation of carboxylic acid groups. As shown in Figure S2 of supplemental information, the formation of carboxyl impurities were avoided by the lack of significant amount of oxygen in our system which typically was operated at $10^{-9}$-$10^{-10}$ torr. This absence of carboxylic acid groups was particularly important for our case in order to reduce $Fe^{3+}$ absorption due to carboxylate salts in the irradiated Nafion areas.

The loss of sulfonate and other groups during irradiation was expected to render the surface less hydrophilic. This change was further supported by the contact angle measurements. Data were summarized as a function of electron doses in Figure 3, showing an apparent jump in the water contact angles in agreement with the FTIR results. Since wetting responses are believed to come from the very top surfaces (~0.5 nm)[39], these increased contact angle values further confirmed the distinct changes in Nafion's surface chemistry.

The immersion of these electron-beam patterned substrates into the $FeCl_3$ solutions caused differential absorption of $Fe^{3+}$ ions according to $SO_3^-$ content. Excessive $FeCl_3$ were removed from these substrates with an aqueous acidified (pH<4) wash before the samples were exposed onto a basic (pH=12.7) DMF wash. Residual surface water content as well as moisture impurities in DMF caused the adsorbed $Fe^{3+}$ ions to precipitate as hydroxides onto these patterned Nafion substrates. Figure 4a provides an iron LMM Auger elemental map collected at a peak of 598 eV. More irons were found in the brighter areas, which were those prevented from irradiation by the mask. Darker regions (squares) were iron-deficient due to the removal of the ionic groups of Nafion that expected to strongly absorb $Fe^{3+}$ ions. Figure 4b depicts the corresponding line scan at the same peak energy. Oscillation peaks due to the preferential distribution of iron concentration can be clearly observed. This selective metal-functionalization is the key step to the formation of patterned SWNT forests[29].

To determine the nature of these iron deposits, TEM investigation (Figure 5) was performed. These precipitates appear to have crystalline morphology with size of c.a. 100 nm in length and c.a. 20 nm in diameter. Selected-area electron diffraction (SAED) investigation further revealed that these crystallites were mixtures of FeO(OH)/FeOCl (see Table S1 in supplemental information). Based on the Fe/Cl ratio (>10) calculated from Auger spectrum (Figure S3), it was deduced that the main component of these crystallites was FeO(OH). The fact that both adventitious carbon and Nafion contribute to the C signal in the Auger spectra and adventitious carbon could have different adsorption rate on irradiated and unirradiated surfaces makes it difficult to estimate the iron surface coverage by comparing the Fe/C peak ratio. The much higher C intensity in Figure S3b and similar peak shape to that in Figure S1 did indicate that part of the irradiated Nafion regions was exposed and devoid of iron deposits.

In order to compare quantitatively the deposition of FeO(OH)/FeOCl crystallites between exposed and unexposed areas, AFM topological imaging was performed. Figure 6a showed the non-irradiation regions were fully covered with FeO(OH)/FeOCl crystallites forming continuous multi-layers (≥2). In contrast, in the e-beam exposed areas the FeO(OH)/FeOCl existed as discrete crystallites in partial monolayer configuration (c.a. 42% coverage in Figure 6b). The surface roughness RMS of these FeO(OH)/FeOCl deposits on the e-beam unexposed regions, as measured by AFM, was found in the order of 10 nm (see Figure 7). Higher RMS values were observed for the irradiated regions, since FeO(OH)/FeOCl crystallites were separated from each other resulting in a rougher surface topology. The plateau in Figure 7 indicated that a steady state condition in FeO(OH)/FeOCl/Nafion roughness has been approached at similar exposure values as those observed from the IR investigation (Figure 2) and contact angle measurements (Figure 3). Much lower electron doses (50, 150μC/cm$^2$) were also tried and the obtained patterns were not well defined (data not shown).

Determining whether these FeO(OH)/FeOCl crystallites are on the top of the exposed Nafion or underneath could prove important to better understand the subsequent forest self-organization of SWNTs onto the irradiated Nafion regions. Contact angle measurements were assessed, but the small differences between irradiated (32±2°) and unirradiated (27±1°) regions were unable to provide a definite conclusion on the placement of these crystallites on the irradiated Nafion regions. The presence of strong Fe Auger signals (as shown in Figure S3 of supplemental information), however, suggested that these FeO(OH)/FeOCl crystallites were placed on top of the irradiated Nafion surface. Thus the question arises on where the $Fe^{3+}$ ions are "stored" in these irradiated Nafion thin films and how they diffuse to the surface to form these FeO(OH)/FeOCl crystallites. The 4-8 nm thickness of the irradiated Nafion films certainly provides adequate porosity and although after irradiation Nafion films become more hydrophobic (contact angle of 31-42°), it is still hydrophilic enough to ensure passage of $Fe^{3+}$ ions. Figure S4 in supplemental information depicts the S 2p (164.40 eV) of thin Nafion films (c.a. 10 nm thick) before and after irradiation (electron dose 1250 µC/cm$^2$). Although Nafion's low S content results in signal to noise ratio far from ideal, we can safely conclude that complete $SO_3^-$ elimination has not occurred and c.a. 47% of sulfonate groups remain present as measured from the S peak areas in Figure S4. In addition, the underlying silanol groups (Si-OH) of native Si oxide layer is also capable of absorbing $Fe^{3+}$ ions[50-52] and their affinity to $Fe^{3+}$ ions could be further enhanced upon e-beam exposure which we have confirmed by other experiments[53]. Upon washing in basic DMF (pH 12.7), Nafion could be plasticized by DMF enabling $Fe^{3+}$ to diffuse out. Residual water content in Nafion as well as moisture impurities in DMF would facilitate $Fe^{3+}$ ions to transform into hydroxides. The Nafion/DMF boundary is the natural interface that the two reagents ($Fe^{3+}$ and $OH^-/H_2O$) would meet and grow crystals of high order due to slow diffusion of both reagents from either side.

Upon immersion of these FeO(OH)/FeOCl decorated Nafion substrates into DMF dispersed shortened-SWNTs acid-base neutralization between the carboxylic acid terminated nanotubes and the basic iron hydroxides provides the initial driving force for SWNT assembly.[29] The strong hydrophobic interaction between adjacent SWNT side walls is believed to further facilitate the bundling along the lateral direction leading to rope-lattice SWNT forest assemblies. Figure 8 illustrates a typical AFM image of the resulting SWNT forests onto grid-patterned Nafion substrates. The majority of these forests were localized within the unexposed FeO(OH)/FeOCl regions although some SWNTs still can be observed in the exposed region. Raman scattering provides further evidence for the preferential self-assembly of SWNT on patterned substrates as shown in Figure 9 where much higher signal intensity in the non-irradiation regions was observed at the characteristic G band (1592 cm$^{-1}$) than that in the exposed areas (peak height ratio $I_a/I_b$=24.6). This ratio is significantly higher than that based on the FeO(OH)/FeOCl surface coverage between the unexposed and e-beam exposed regions, which was estimated to a ratio of c.a. 2.4 according to AFM measurement. Since nanotube self-assembly is based on electrostatic ineractions[29] and such interactions extend for a few to tens of nanometers[54], the positive polarity of the discrete FeO(OH)/FeOCl crystallites within exposed regions could be counteracted by the residual negative polarity of Nafion or Si substrates, thus reducing the electrostatic attraction of the negatively charged SWNTs. In contrast, fully covered multilayer FeO(OH)/FeOCl in the unexposed domains assured a clearly surface charge that attracted significantly large number of SWNTs. Currently, efforts are underway to improve SWNT forest patterning and these will be presented in subsequent publications.

## 4. Conclusions

In this study we have demonstrated that low-energy (500 eV) electron-beam could sufficiently pattern Nafion and assist in selective localization of SWNT forests. During electron-beam exposure, a large number of sulfonate groups of Nafion were cleaved and removed in ultra-high vacuum, thereby reducing the ability of irradiated regions withhold $Fe^{3+}$ cations. Subsequent immersion to basic DMF solution resulted in the preferential deposition of iron hydroxides onto non-irradiation areas which was confirmed by Auger chemical mapping and AFM. This provided the basis for selective deposition of SWNTs into forest arrays as demonstrated by AFM and Raman spectra. These aligned and patterned SWNT forests could find a number of applications, such as field emission electron sources and biomaterial/nanotube hybrids for biosensor arrays.

**Acknowledgement.** The authors thank S. Kim for helpful discussion and assistance with the liquid drop contact angle measurement, and R. Li for his help in the TEM experiments. The authors gratefully acknowledge the financial support of U.S. Army Research Office (grant # ARO-DAAD-19-02-1-0381).

**Supporting Information Available:** Indexing table of FeO(OH)/FeOCl TEM diffraction pattern for Figure 5b (Table S1); Auger spectrum of the dipping acquired Nafion films (Figure S1); IR-ATR spectra (1500-1900 $cm^{-1}$) of the irradiated Nafion 112 membranes (Figure S2); Auger spectra of FeO(OH)/FeOCl crystallites on irradiated and unirradiated Nafion films (Figure S3); XPS investigation on S 2p of exposed and unexposed Nafion films (Figure S4) (PDF). This material is available free of charge via the Internet at http://pubs.acs.org.

# Supporting information

# Preferential Forest-assembly of Single-Wall Carbon Nanotubes on Low-energy Electron-beam Patterned Nafion Films

*Haoyan Wei‡, Sang Nyon Kim†, Harris L. Marcus‡\*, and Fotios Papadimitrakopoulos†\**

‡ Department of Materials Science and Engineering, Institute of Materials Science, University of Connecticut, Storrs, CT 06269, USA

† Nanomaterials Optoelectronics Laboratory, Polymer Program, Institute of Materials Science, Department of Chemistry, University of Connecticut, Storrs, CT 06269, USA

Email: papadim@mail.ims.uconn.edu, hmarcus@mail.ims.uconn.edu.

**Table S1.** Experimental electron diffraction spacings for FeO(OH) and FeOCl crystals in comparison with Joint Committee on Powder Diffraction Standards/International Center for Diffraction Data (JCPDS/ICDD) powder diffraction file (PDF) data.

| | | | | JCPDS/ICDD PDF data | | | | | |
| | Experimental data | | | FeOCl | | | FeO(OH) | | |
| n | $D_n$(mm) | $D_n/D_1$ | $D_n/D_2$ | d (Å) | $d_1/d_n$ | hkl | d′ (Å) | $d'_n/d'_1$ | h′k′l′ |
|---|---|---|---|---|---|---|---|---|---|
| 1 | 8.5 | 1.000 | | $d_1$=8.00000 | 1.000 | 010 | | | |
| 2 | 12 | 1.412 | 1.000 | | | | $d'_1$=6.26300 | 1.000 | 020 |
| 3 | 18.8 | 2.212 | 1.567 | $d_2$=3.41903 | 2.340 | 110 | | | |
| 4 | 24 | 2.8235 | 2.000 | | | | $d'_2$=3.28996 | 1.904 | 120 |
| 5 | 26.8 | 3.153 | 2.233 | $d_3$=2.52996 | 3.162 | 021 | $d'_3$=2.97027 | 2.1085 | 011 |
| 6 | 29.5 | 3.4705 | 2.458 | $d_4$=2.35945 | 3.391 | 111 | $d'_4$=2.47045 | 2.535 | 031 |
| 7 | 31.5 | 3.706 | 2.625 | $d_5$=2.05005 | 3.902 | 031 | $d'_5$=2.35945 | 2.654 | 111 |
| 8 | 35.5 | 4.176 | 2.958 | $d_6$=1.89017 | 4.232 | 200 | $d'_6$=2.08980 | 3.000 | 060 |
| 9 | 38 | 4.4705 | 3.167 | $d_7$=1.80999 | 4.420 | 131 | $d'_7$=1.93664 | 3.234 | 200 |
| 10 | 41 | 4.8235 | 3.417 | $d_8$=1.63971 | 4.879 | 002 | $d'_8$=1.84838 | 3.388 | 220 |

The ring patterns in Figure 5b were indexed using the fundamental relationship (Rd=λL) in diffraction patterns by calculating the ratios of the measured ring diameters against a table of ratios of the interplanar spacings for the crystals of interest. D is the measured diameters of diffraction rings, R=D/2, d and d′ are the interplanar spacing for FeOCl and FeO(OH) respectively, λ is the wavelength, L is the camera length, *hkl* and *h′k′l′* are the Miller indices of planes in FeOCl and FeO(OH) crystals respectively.

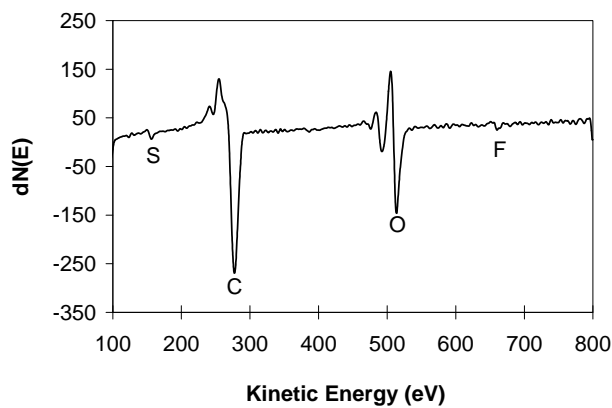
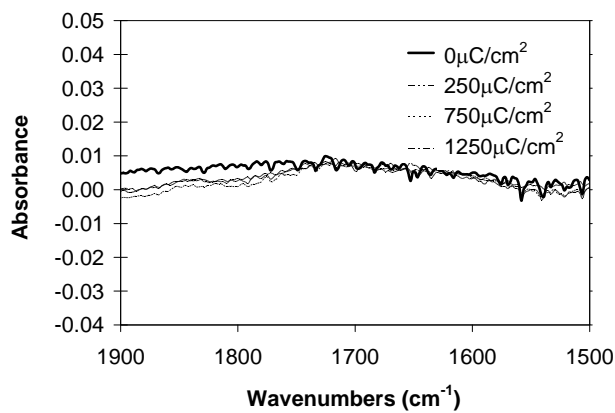

**Figure S1.** Auger spectrum indicating the successful deposition of Nafion on Si substrates (see text for details).

**Figure S2.** IR-ATR spectra of Nafion 112 membranes irradiated at 3 keV with different doses. The absence of any significant absorbance in the 1500-1900 cm$^{-1}$ range indicated that carboxyl functionalities (typically reported at 1770 cm$^{-1}$) did not take place.

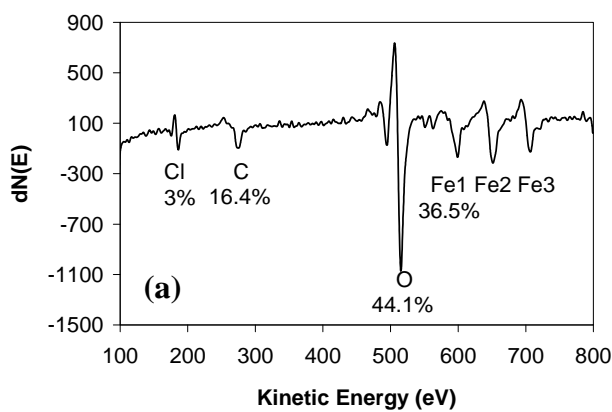
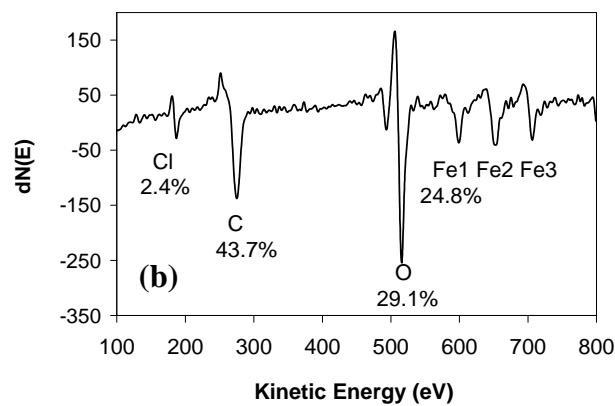

**Figure S3.** Auger investigation on iron deposits in the unirradiated (a) and the irradiated (500 eV, 1250 μC/cm$^2$) regions (b) of Nafion.

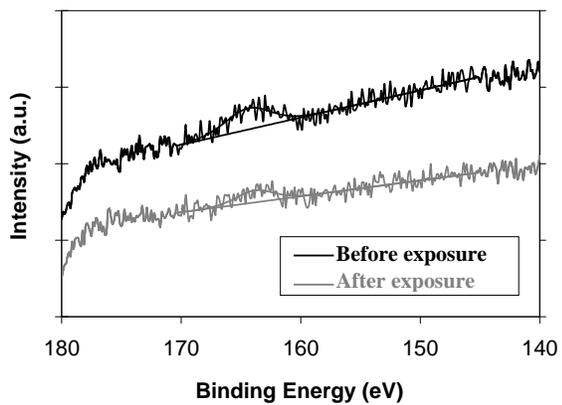

**Figure S4.** After Nafion was exposed to electron-beam (500 eV, 1250 μC/cm$^2$) XPS showed a 53% intensity reduction of S 2p signal (peak position 164.40 eV, FWHM 4.452 eV).

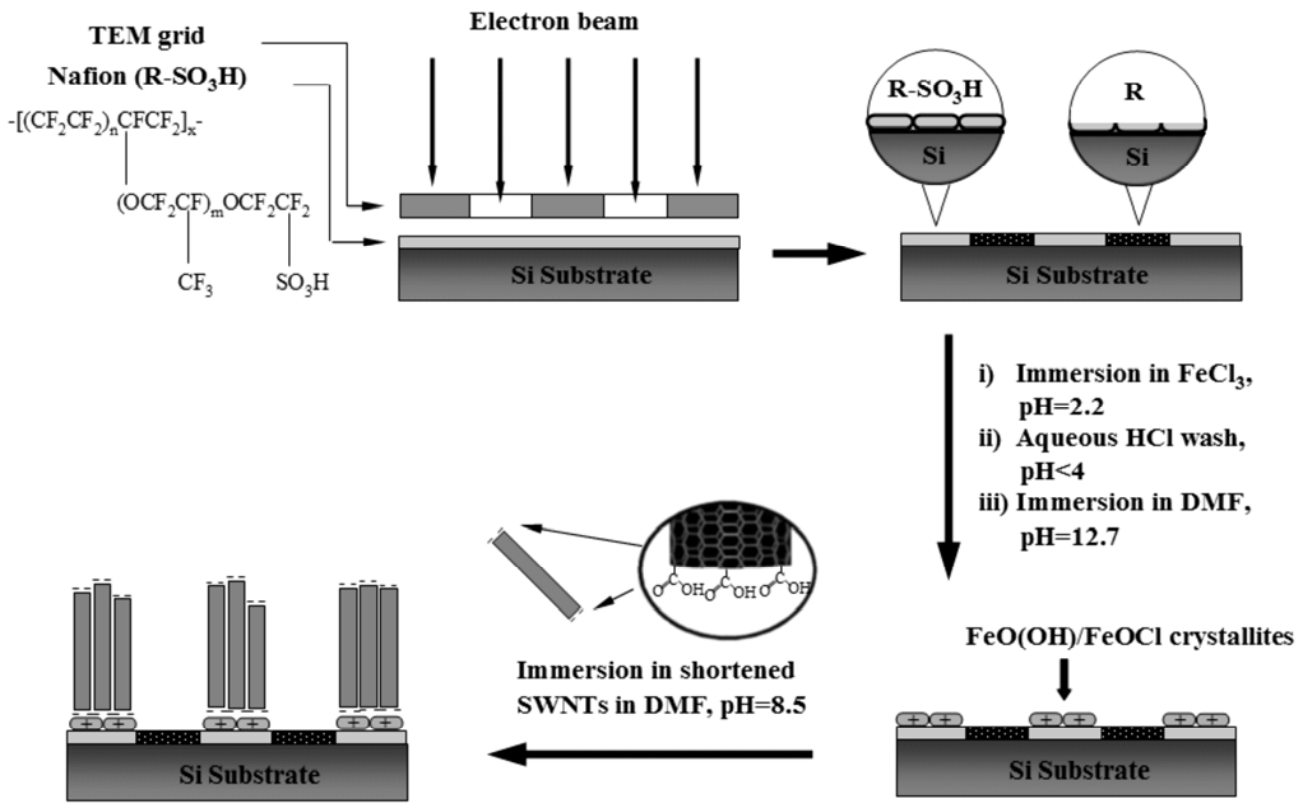

**Scheme 1**. Schematic representation of the formation of patterned SWNT forests.

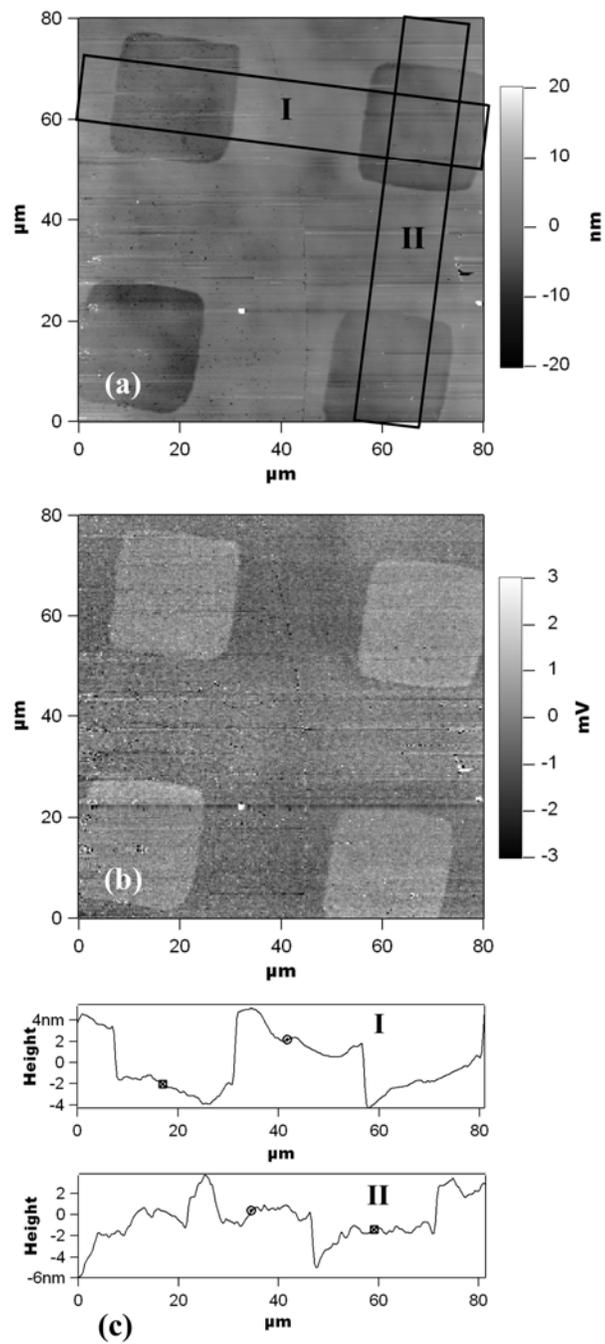

**Figure 1.** AFM images of grid-patterned Nafion irradiated with 500 eV e-beam and dose of 750 µC/cm². Squares were the exposed regions. Topological (a) and lateral force microscopy (b) images. (c) Topological profiles along the indicated parallelograms in (a).

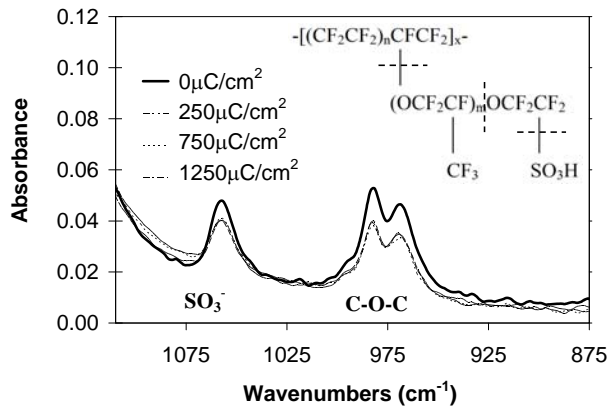

**Figure 2.** IR-ATR spectra of Nafion 112 membranes irradiated at 3 keV with different doses. Top right inset shows the chemical structure of Nafion and the likely scission sites indicated by dash lines[46].

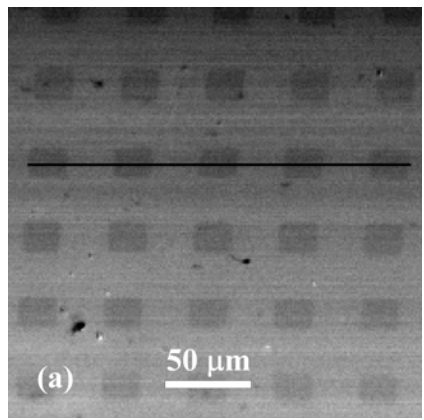

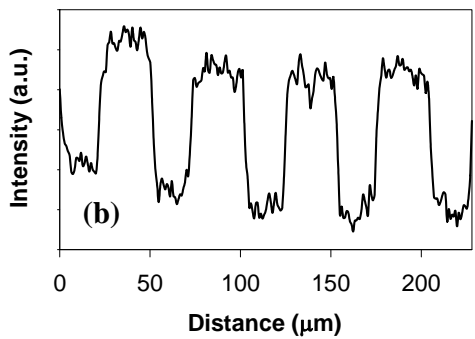

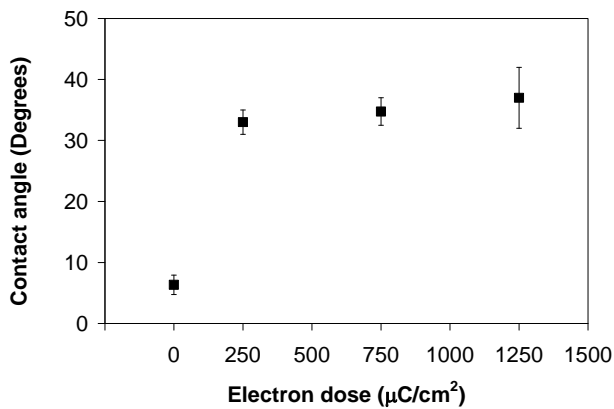

**Figure 3.** Contact angles of Nafion irradiated at 500 eV with different doses.

**Figure 4.** Typical Auger survey of iron deposits of grid-patterned Nafion at 500 eV with electron dose of 750 $\mu C/cm^2$. Squares were exposed regions. Auger mapping of elemental Fe LMM at peak energy 598 eV (a). Corresponding line scan (see horizontal black line in (a)) of Fe as a function of distance (b).

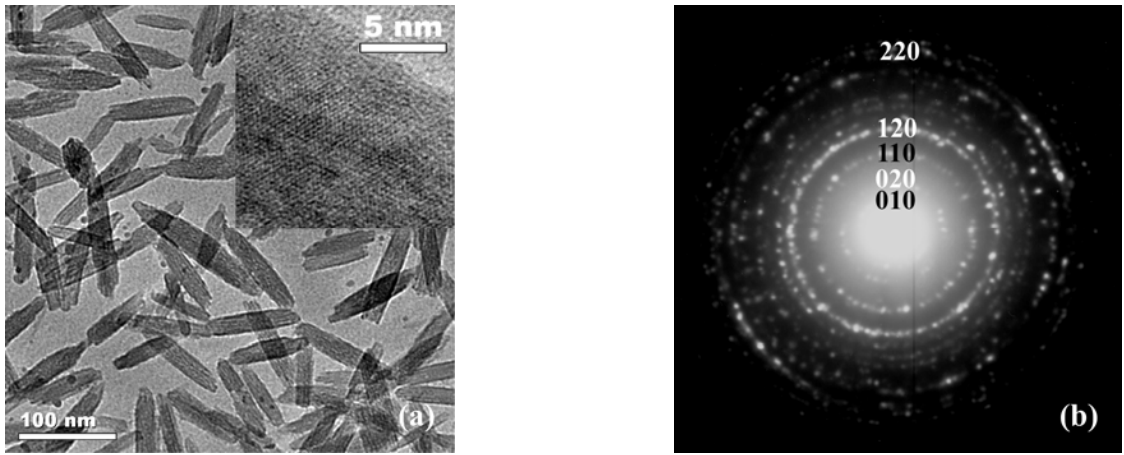

**Figure 5.** TEM bright field (BF) image (a) and diffraction patterns (b) of FeO(OH)/FeOCl precipitates. Inset in (a) shows a typical HRTEM lattice image of the crystalline FeO(OH)/FeOCl precipitates. White and dark *hkl* annotations in (b) correspond to the expected FeO(OH)'s and FeOCl's indices respectively.

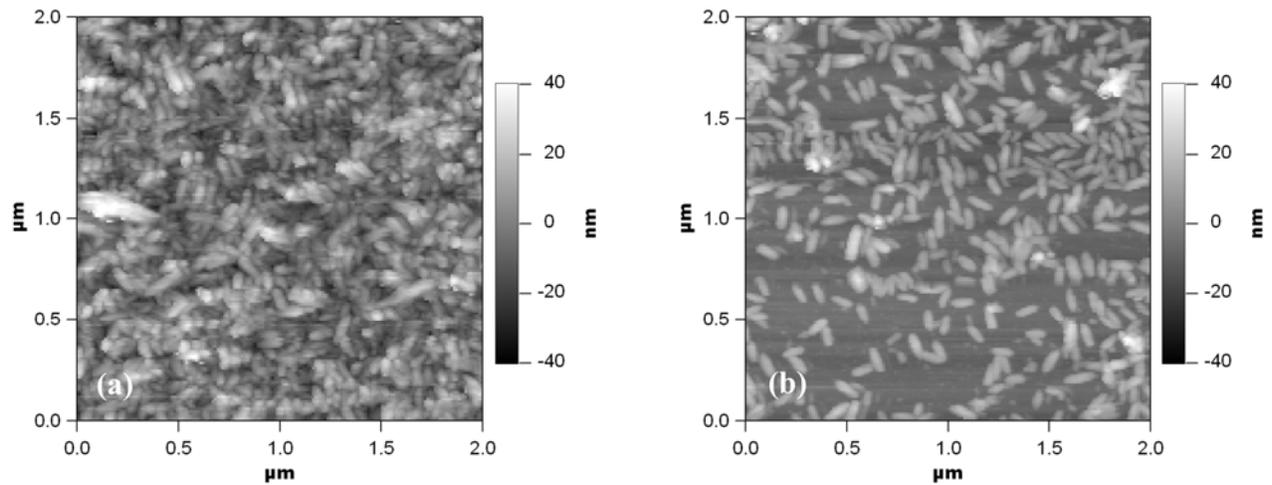

**Figure 6.** AFM images show the FeO(OH)/FeOCl crystallites exist as continuous multi-layers on non-irradiation regions (a) while in partial monolayer configuration on irradiation regions (500 eV, 750 µC/cm$^2$) (b).

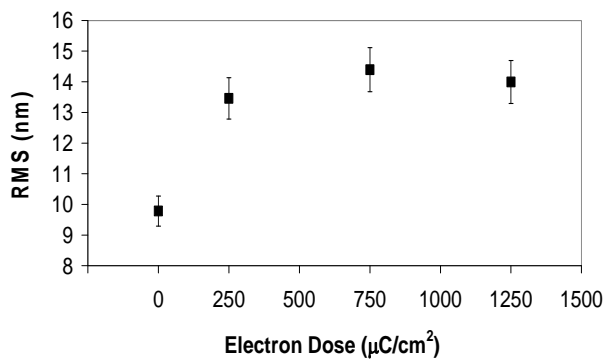

**Figure 7.** AFM measured surface roughness of FeO(OH)/FeOCl deposits on 500 eV electron-beam exposed Nafion with different doses.

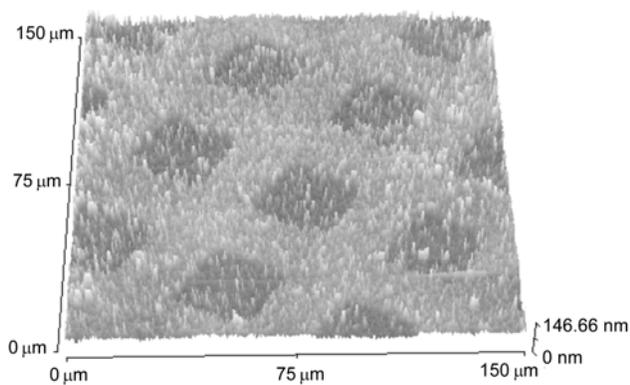

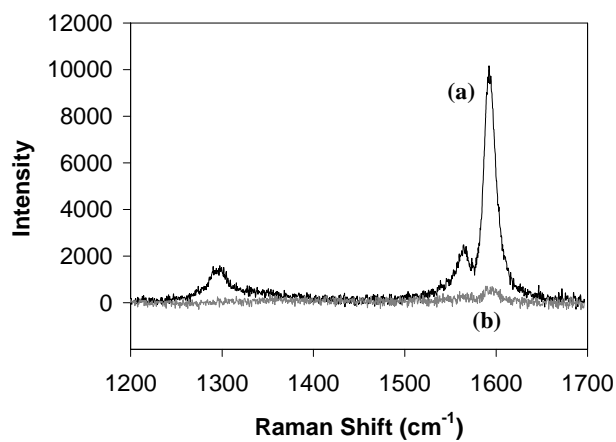

**Figure 8.** Representative AFM image of patterned SWNT forests. Nafion was patterned with 500 eV e-beam irradiation (1250 $\mu C/cm^2$).

**Figure 9.** 785 nm Raman spectra of the patterned SWNT forests taken at unexposed region (a) and exposed region (b) respectively. Nafion was patterned with 500 eV e-beam irradiation (1250 $\mu C/cm^2$).

Table of Contents Synopsis

With the aid of low-energy electron-beam (500 eV) writing, patterns of perpendicularly-aligned Single-wall carbon nanotube forest were realized on Nafion-irradiated substrates with the help of $Fe^{3+}$-assisted self-assembly.

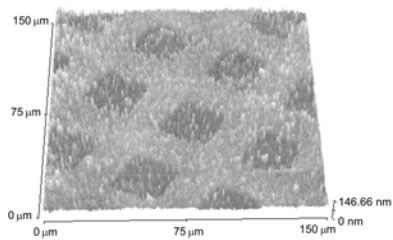